\documentclass[twocolumn]{aastex631}
\usepackage{mathrsfs}
\shorttitle{Solar-like dynamos and rotational scaling of cycles}
\shortauthors{K\"apyl\"a, P.J.}
\newcommand{\fff}{{\bm f}}

\newcommand{\uuu}{{\bm u}}

\newcommand{\xxx}{{\bm x}}

\newcommand{\ooo}{{\bm \omega}}
\newcommand{\AAA}{{\bm A}}

\newcommand{\BBB}{{\bm B}}

\newcommand{\FFF}{{\bm F}}

\newcommand{\JJJ}{{\bm J}}

\newcommand{\UUU}{{\bm U}}

\newcommand{\gggg}{{\bm g}}

\newcommand{\mUUr}{\overline{U}_r}

\newcommand{\mUUt}{\overline{U}_\theta}
\newcommand{\mUUp}{\overline{U}_\phi}

\newcommand{\mBBr}{\overline{B}_r}

\newcommand{\mBBt}{\overline{B}_\theta}
\newcommand{\mBBp}{\overline{B}_\phi}

\newcommand{\EQ}{\begin{equation}}
\newcommand{\EN}{\end{equation}}
\newcommand{\EQA}{\begin{eqnarray}}
\newcommand{\ENA}{\end{eqnarray}}
\newcommand{\brac}[1]{\langle #1 \rangle}
\newcommand{\pd}{\partial}

\newcommand{\mean}[1]{\overline{#1}}

\newcommand{\cP}{c_{\rm P}}
\newcommand{\cV}{c_{\rm V}}

\newcommand{\urms}{u_{\rm rms}}

\newcommand{\orms}{\omega_{\rm rms}}

\newcommand{\Pcyc}{P_{\rm cyc}}
\newcommand{\Prot}{P_{\rm rot}}
\newcommand{\Beq}{B_{\rm eq}}

\newcommand{\chiSGS}{\chi_{\rm SGS}}

\newcommand{\Co}{{\rm Co}}

\newcommand{\Pe}{{\rm Pe}}

\newcommand{\PraSGS}{{\rm Pr}_{\rm SGS}}

\newcommand{\PrM}{{\rm Pr}_{\rm M}}

\newcommand{\Rey}{{\rm Re}}

\newcommand{\ReM}{{\rm Re}_{\rm M}}
\newcommand{\Ro}{{\rm Ro}}

\newcommand{\taucool}{\tau_{\rm cool}}

\newcommand{\mOm}{\mean{\Omega}}

{}

\newcommand{\Rgas}{{\cal R}}
\newcommand{\FFFrads}{{\bm F}_{\rm rad}}
\newcommand{\FFFSGSs}{{\bm F}_{\rm SGS}}

\newcommand{\Ekin}{E_{\rm kin}}
\newcommand{\Emag}{E_{\rm mag}}

\def\onethird{{\textstyle{1\over3}}}

\def\onehalf{{\textstyle{1\over2}}}

\newcommand{\Fig}[1]{Figure~\ref{#1}} 
\newcommand{\Figu}[1]{Figure~\ref{#1}}

\newcommand{\Table}[1]{Table~\ref{#1}}

\newcommand{\K}{\,{\rm K}}
\newcommand{\s}{\,{\rm s}}
\newcommand{\m}{\,{\rm m}}

\graphicspath{{./}{fig/}}
\usepackage{bm}

\begin{document}

\title{Solar-like dynamos and rotational scaling of cycles from
  star-in-a-box simulations}

\correspondingauthor{Petri J. K\"apyl\"a}
\email{pkaepyl@uni-goettingen.de}

\author[0000-0002-0786-7307]{Petri J. K\"apyl\"a}
\affiliation{Georg-August-Universit\"at G\"ottingen\\
Institute for Astrophysics and Geophysics\\
Friedrich-Hund-Platz 1 \\
37077 G\"ottingen, Germany}

\begin{abstract}

Magnetohydrodynamic star-in-a-box simulations of convection and
dynamos in a solar-like star with different rotation rates are
presented. These simulations produce solar-like differential rotation
with a fast equator and slow poles, and magnetic activity that
resembles that of the Sun with equatorward migrating activity at the
surface. Furthermore, the ratio of rotation to cycle period is almost
constant as the rotation period decreases in the limited sample
considered here. This is reminiscent of the suggested inactive branch
of stars from observations and differs from most earlier simulation
results from spherical shell models. While the exact excitation
mechanism of the dynamos in the current simulations is not yet clear,
it is plausible that the greater freedom that the magnetic field has
due to the inclusion of the radiative core and regions exterior to the
star are important in shaping the dynamo.

\end{abstract}


\section{Introduction} \label{sec:intro}

The Sun maintains a global dynamo with a magnetic cycle of
approximately 22 years, with activity appearing at the surface at
midlatitudes and propagating equatorward as the cycle progresses
\citep[e.g.][]{2010LRSP....7....1H}. Three-dimensional
magnetohydrodynamic (MHD) simulations struggle to reproduce such
cycles: often the activity propagates poleward
\citep[e.g.][]{BMBBT11,NBBMT13}, active latitudes do not coincide with
those in the Sun \citep[e.g.][]{GCS10}, or there is a mismatch between
the simulated and solar cycle periods
\citep[e.g.][]{KMB12,2018A&A...616A..72W}. Furthermore, the
simulations usually require substantially faster rotation than in the
Sun to achieve cyclic dynamos \citep[e.g.][]{2018A&A...616A.160V}.
Another issue arises when simulations at different rotation rates are
confronted with observations: cycles observed from stars other than
the Sun suggest that in the vicinity of the solar Rossby number, which
is the ratio of the rotation period to convective turnover time, the
ratio of rotation to cycle period is increasing as the Rossby number
decreases \citep[e.g.][]{2018A&A...619A...6O}. Simulations often
produce the opposite trend
\citep[e.g.][]{2018A&A...616A..72W,2018ApJ...863...35S}. Typically
more than one of these defects is found in any given simulation.

There are several possible reasons for the mismatch between
simulations and reality. A major factor in this is likely to be the
inability of current simulations to capture stellar convective flows
accurately enough. This is known as the \emph{convective conundrum},
and manifested by too high large-scale velocity amplitudes in
simulations in comparison to the Sun
\citep[e.g.][]{HDS12,2016AdSpR..58.1475O,2020RvMP...92d1001S}. This
often leads to anti-solar differential rotation in simulations with
solar luminosity and rotation rate \citep[e.g.][]{KKB14}, and very
high resolutions in addition to highly supercritical dynamos are
likely needed to overcome this \citep[][]{2021NatAs...5.1100H}.

Another factor is that stellar dynamo simulations are often done in
spherical shells where only the convection zone (CZ), or additionally
a part of the radiative core is modeled
\citep[e.g.][]{2019ApJ...880....6G,2020ApJ...893..107B}. This
necessitates the use of boundary conditions which may not always be
appropriate and which can affect the dynamo solutions in ways that are
\emph{a priori} not obvious \citep[e.g.][]{CBKK16}. In the present
study a star-in-a-box model, where a spherical star is embedded into a
Cartesian cube, is used to model a solar-like star in a rotational
regime where cyclic solutions are excited. The model thus includes the
radiative core and regions exterior to the star. The former enables
contributions to the dynamo from the interface between radiative and
convective zones, whereas the latter is usually not considered to be
important in the maintenance of the dynamo. However, including the
outer layers is less restrictive than imposing mathematically or
numerically convenient boundary conditions on the flows and magnetic
fields, and which can also affect the resulting dynamo solution
\citep[e.g.][]{WKKB16}.

\section{Model} \label{sec:model}

The star-in-a-box model described in \cite{2021A&A...651A..66K} is
used; see also \cite{DSB06}. A star of radius $R$ is embedded into a
Cartesian cube with a side length $H=2.2R$. Equations governing the
system are:
\begin{eqnarray}
\frac{\pd\AAA}{\pd t}&=&\UUU\times\BBB-\eta\mu_0\JJJ,\label{equ:indu} \\
\frac{D\ln\rho}{Dt}&=&-\bm\nabla\bm\cdot\UUU,\label{equ:conti} \\
\frac{D\uuu}{Dt}&=&-\bm\nabla\Phi-\frac{1}{\rho}(\bm\nabla p-\bm\nabla\bm\cdot2\nu\rho\bm{\mathsf{S}}-\JJJ\times\BBB)\nonumber \\ && \hspace{3.5cm}-2\bm\Omega\times\UUU+\fff_d,\label{equ:mom} \\
\rho T\frac{Ds}{Dt}&=&-\bm\nabla\!\bm\cdot\!\left(\FFF_{\rm rad}\!+\!\FFF_{\rm SGS}\right)\!+\!{\mathscr H}\!\!-\!\!{\mathscr C}\!+\!2\nu\bm{\mathsf{S}}^2\!\!+\!\mu_0\eta\JJJ^2\!\!,\label{equ:entro}
\end{eqnarray}
where $\AAA$ is the magnetic vector potential, $\UUU$ is the velocity,
$\BBB=\bm\nabla\times\AAA$ is the magnetic field,
$\JJJ=\bm\nabla\times\BBB/\mu_0$ is the current density, $\mu_0$ is
the permeability of vacuum, $\eta$ is the magnetic diffusivity, $D/Dt
= \pd/\pd t+\UUU\bm\cdot\bm\nabla$ is the advective derivative, $\rho$
is the fluid density, $\Phi$ is the gravitational potential, $p$ is
the pressure, and $\nu$ is the kinematic viscosity. The traceless
rate-of-strain tensor is given by
$\mathsf{S}_{ij}=\onehalf\left(\frac{\pd U_i}{\pd x_j}+\frac{\pd
  U_j}{\pd x_i}\right)-\onethird\delta_{ij}\bm\nabla\bm\cdot\UUU$,
where $\delta_{ij}$ is the Kronecker delta. The angular velocity is
given by $\bm\Omega=(0,0,\Omega_0)$, $\fff_d$ is a damping function,
$T$ is the temperature, and $s$ is the specific entropy. $\FFFrads$
and $\FFFSGSs$ are the radiative and subgrid-scale (SGS) entropy
fluxes, and ${\mathscr H}$ and ${\mathscr C}$ describe heating and
cooling, respectively.

The gas obeys an ideal gas equation of state with $p=\Rgas\rho T$,
where $\Rgas=\cP-\cV$ is the gas constant, and $\cP$ and $\cV$ are the
heat capacities at constant pressure and volume, respectively. The
gravitational potential $\Phi$ corresponds to a polytrope of index
$n=1.5$ of a main-sequence M5 star \citep[see Appendix~A of][]{DSB06}.
Flows in the exterior to the star are damped through the term
$\fff_d=-\frac{\UUU}{\tau_{\rm damp}}f_{\rm e}(r)$, where $\tau_{\rm
  damp}$ is a damping timescale, and $f_{\rm
  e}(r)=\frac{1}{2}\left(1+\tanh\frac{r-r_{\rm damp}}{w_{\rm
    damp}}\right)$, where $r_{\rm damp}=1.03R$ and $w_{\rm
  damp}=0.03R$. The damping timescale $\tau_{\rm
  damp}\approx0.2\tau_{\rm ff}$, where $\tau_{\rm ff} = \sqrt{R^3/GM}$
is the free-fall time, $G$ is the gravitational constant, and $M$ is
the mass of the star.

The radiative flux is given by $\FFFrads=-K\bm\nabla T$, where
\begin{eqnarray}
K(\rho,T)=K_0(\rho/\rho_0)^{a-1}(T/T_0)^{b+3},
\end{eqnarray}
with $a=-1$ and $b=7/2$ corresponds to the Kramers opacity law
\citep[e.g.][]{2000gac..conf...85B}. Additional SGS entropy flux is
included with $\FFFSGSs=-\chiSGS\rho\bm\nabla s'$, where
$s'=s-\brac{s}_t$, are the fluctuations of the entropy, and
$\brac{s}_t(\xxx,t)$ is a running temporal mean computed over an
interval of ten free-fall times. Nuclear energy production in the core
of the star is parameterized by the heating term ${\mathscr H}$ with a
Gaussian profile, ${\mathscr H}(r)=\frac{L_{\rm sim}}{(2\pi
  w_L^2)^{3/2}}\exp\left(-\frac{r^2}{2w^2_L} \right)$, where $L_{\rm
  sim}$ is the luminosity and $w_L=0.162R$ is the width of the
Gaussian. The cooling term ${\mathscr C}$ models radiative losses
above the stellar surface with ${\mathscr
  C}(\xxx)=\rho\cP\frac{T(\xxx)-T_{\rm surf}}{\taucool}f_{\rm e}(r)$,
where $\taucool=\tau_{\rm damp}$ is a cooling timescale and $T_{\rm
  surf}$ the fixed surface temperature.

The fluid and magnetic Reynolds numbers, and the P\'eclet number are
given by $\Rey=\urms/(\nu k_1)$, $\ReM=\PrM\Rey=\urms/(\eta k_1)$,
$\Pe=\PraSGS\Rey=\urms/(\chiSGS k_1)$, where $\urms$ is the
volume-averaged rms-velocity in the CZ and $k_1=2\pi/\Delta R$ the
wavenumber corresponding to the approximate depth of the CZ with
$\Delta r=0.35R$, and where $\PrM=\nu/\eta$ and $\PraSGS=\nu/\chiSGS$
are the magnetic and SGS Prandtl numbers, respectively. Rotational
influence on the flow is measured by the Coriolis number,
$\Co=2\Omega_0/(\urms k_1)$. Magnetic fields are measured in terms of
the equipartition field strength $\Beq=\brac{\sqrt{\mu_0 \rho
    \UUU^2}}$ where $\brac{.}$ refers to volume averaging over the
CZ. Mean values are taken to be azimuthal averages and denoted by
overbars. The {\sc Pencil Code}
\citep{2021JOSS....6.2807P}\footnote{\href{https://github.com/pencil-code/}{https://github.com/pencil-code/}}
was used to make the simulations.

\begin{deluxetable*}{lrrrrcccccccc}
\tablecaption{Summary of the
  simulations. $\Emag=\onehalf\brac{\BBB^2/\mu_0}$ and
  $\Ekin=\onehalf\brac{\rho\UUU^2}$ are the total magnetic and kinetic
  energies, and $\Emag^{\rm pol}=\onehalf
  \brac{(\mBBr^2+\mBBt^2)/\mu_0}$, $\Emag^{\rm tor}=\onehalf
  \brac{\mBBp^2/\mu_0}$, $\Ekin^{\rm MC}=\onehalf
  \brac{\rho(\mUUr^2+\mUUt^2)}$, and $\Ekin^{\rm DR}=\onehalf
  \brac{\rho \mUUp^2}$ refer to the energies of the poloidal and
  toroidal magnetic fields, and the meridional circulation and
  differential rotation, respectively. $P_{\rm cyc}^{\rm surf}$ and
  $P_{\rm cyc}^{\rm deep}$ are the cycle periods measured from $\mBBr$
  at the surface and $\mBBp$ at the base of the CZ. $\PrM=0.5$ and
  $\PraSGS=0.2$ in all runs except run~C1 where $\PrM=1$.
  \label{tab:tab1}}
\tablewidth{700pt} \tabletypesize{\scriptsize}
\tablehead{ \colhead{Run} & \colhead{$\Co$} & \colhead{$\Rey$} &
  \colhead{$\Pe$} & \colhead{$\ReM$} & \colhead{$\Emag/\Ekin$} &
  \colhead{$\Ekin^{\rm MC}$} & \colhead{$\Ekin^{\rm DR}$} &
  \colhead{$\Emag^{\rm pol}$} & \colhead{$\Emag^{\rm tor}$} &
\colhead{$\Prot/P_{\rm cyc}^{\rm surf} [10^{-3}]$} &
\colhead{$\Prot/P_{\rm cyc}^{\rm deep} [10^{-3}]$} &
\colhead{Grid}
} 
\startdata
A  &  5.6 &  55 &  11  &  27  &  0.61  &  0.011  &  0.038  &  0.091  &  0.128  &  $3.7\pm0.8$  &  $3.4\pm0.5$ & $288^3$ \\
B  &  7.0 &  53 &  10  &  26  &  0.63  &  0.010  &  0.050  &  0.094  &  0.152  &  $5.5\pm0.4$  &  $5.1\pm0.3$ & $288^3$ \\
C  & 10.0 &  49 &   9  &  24  &  0.61  &  0.008  &  0.050  &  0.096  &  0.146  &  $4.6\pm0.4$  &  $4.1\pm0.2$ & $288^3$ \\
Cm &  9.7 & 102 &  20  &  51  &  0.66  &  0.008  &  0.034  &  0.055  &  0.081  &  $5.1\pm0.5$  &  $4.9\pm0.4$ & $576^3$ \\
C1 &  9.9 & 100 &  20  & 100  &  0.71  &  0.008  &  0.033  &  0.030  &  0.044  &  $5.0\pm0.8$  &  $5.5\pm0.8$ & $576^3$ \\
Ch &  9.5 & 260 &  52  & 130  &  0.69  &  0.008  &  0.035  &  0.026  &  0.041  &      (3.6)    &    (3.6)     & $1152^3$ \\
D  & 17.3 &  42 &   8  &  21  &  0.63  &  0.006  &  0.019  &  0.117  &  0.071  &  $4.8\pm0.3$  &  $4.0\pm0.3$ & $288^3$
\enddata
\end{deluxetable*}

The set-up of the simulations is otherwise identical to those in
\cite{2021A&A...651A..66K} except that the amplitude of the radiative
conductivity $K_0$ is enhanced such that in the thermodynamically
saturated state the star has a radiative core (CZ) that encompasses
roughly two thirds (one third) of stellar radius. Furthermore, the
diffusion coefficients $\eta$, $\nu$, and $\chiSGS$ have radial
profiles such that their values in the radiative core are $10^2$
smaller than in the CZ to avoid diffusive spreading of magnetic fields
and flows into the core. This nevertheless happens in many runs during
the initial transient toward a statistically steady state for the
flow. This is due to the fact that the initial state of the
simulations is an isentropic polytrope, and because of this choice,
the star is fully convective in the early stages. To circumvent this
issue the magnetic field is rescaled to $10^{-6}\Beq$ level after a
statistically steady state for the flow and thermodynamics is
reached. The simulations are then further evolved until the magnetic
field reaches a statistically steady state. The simulations at the
higher resolutions ($576^3$ and $1152^3$) were remeshed from such
saturated snapshots from the low resolution ($288^3$) cases.

\section{Results} \label{sec:results}

The simulations are summarized in Table~\ref{tab:tab1}. The models
cover a modest range of Coriolis numbers between 5.6 and 17 where
cyclic dynamos with a dominating axisymmetric magnetic fields are
found. Runs with slower rotation produce quasi-static magnetic fields
whereas for more rapid rotation non-axisymmetric fields and less
coherent cycles become dominant. The run with the highest resolution
(Ch) has completed only one full cycle and therefore it is not used in
the statistical analysis of the cycle periods, but merely to
demonstrate that the cycles persists also at higher Reynolds
numbers. A more comprehensive study of the simulations including the
slower and faster rotation cases will be presented elsewhere.

\subsection{Magnetic fields and cycles}

The current simulations produce dynamos where the magnetic energy
$\Emag$ is a significant fraction of the kinetic energy $\Ekin$; see
the sixth column of Table~\ref{tab:tab1}. The ratio $\Emag/\Ekin$ is
practically constant as a function of $\Co$ in the parameter regime
studied here, which differs from the scaling found in
\cite{2019ApJ...876...83A}, and the MAC balance which is often assumed
to hold for the saturation level of magnetic fields
\citep[e.g.][]{2017LRSP...14....4B}.

\begin{figure*}
  \includegraphics[width=0.33\textwidth]{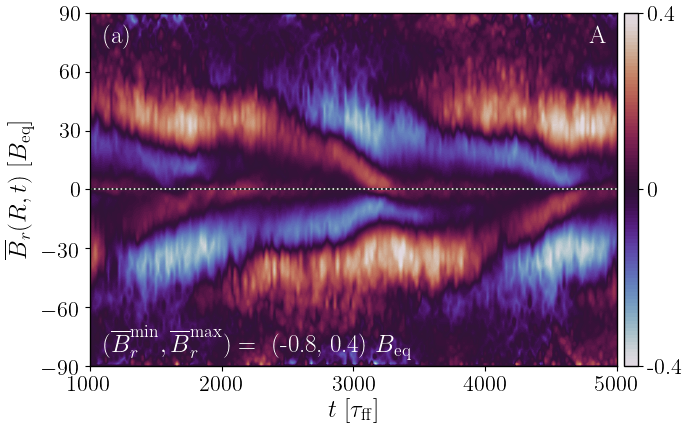}\includegraphics[width=0.33\textwidth]{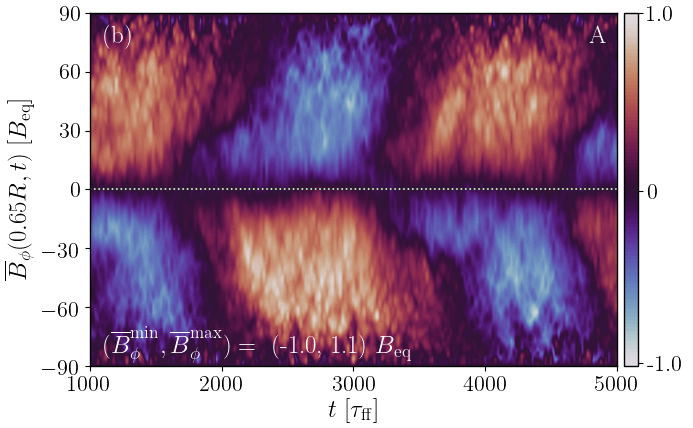}\includegraphics[width=0.32\textwidth]{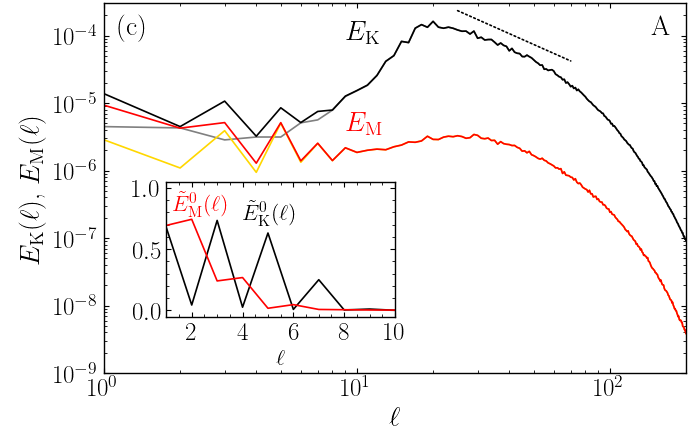}\\
  \includegraphics[width=0.33\textwidth]{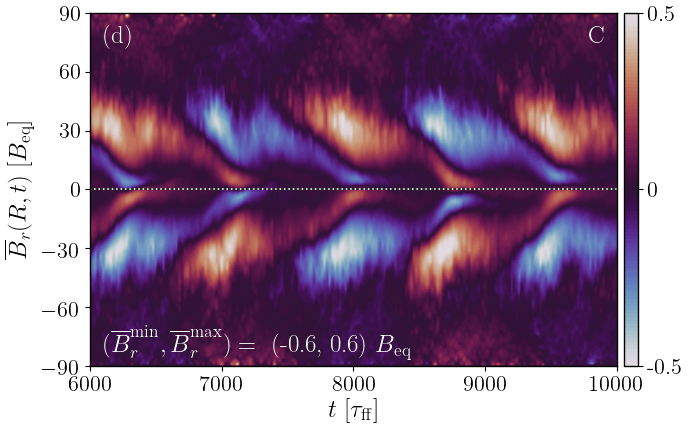}\includegraphics[width=0.33\textwidth]{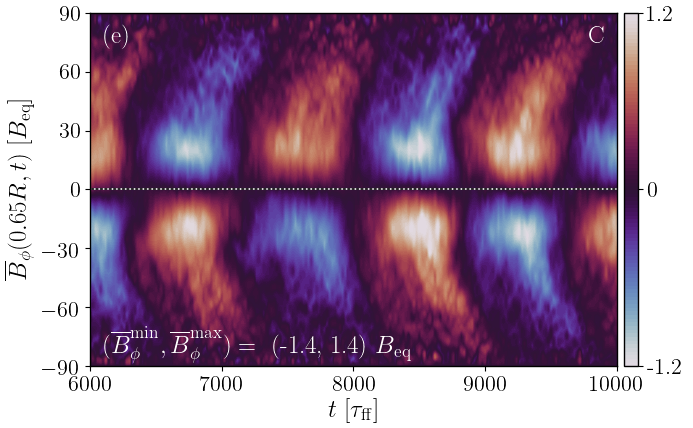}\includegraphics[width=0.32\textwidth]{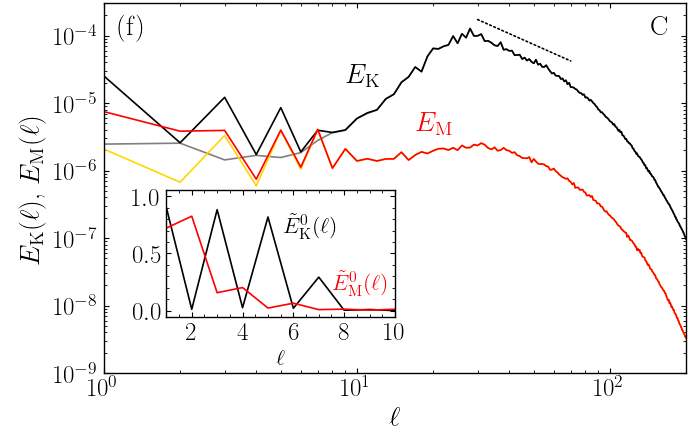}\\
  \includegraphics[width=0.33\textwidth]{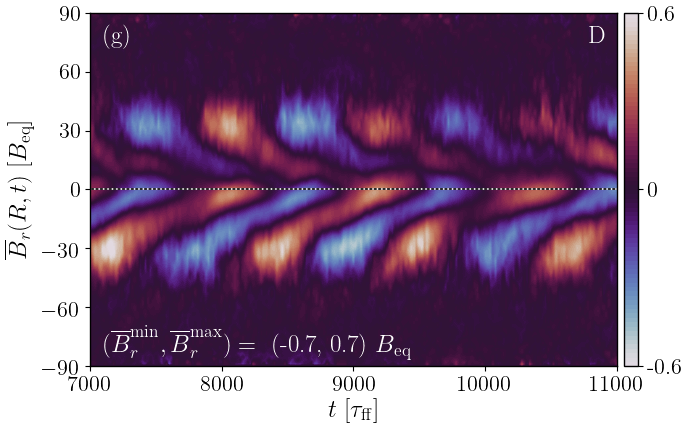}\includegraphics[width=0.33\textwidth]{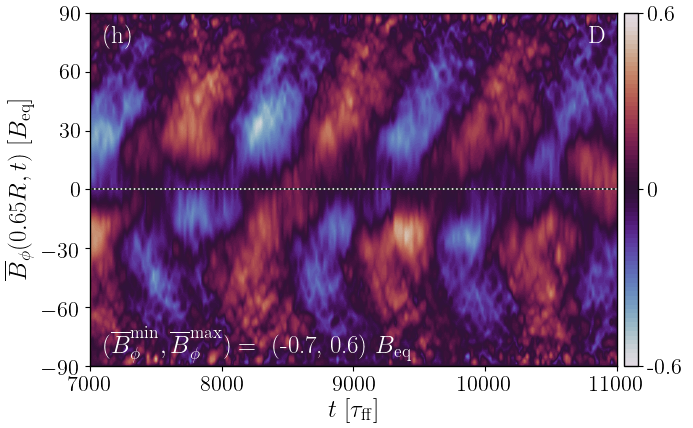}\includegraphics[width=0.32\textwidth]{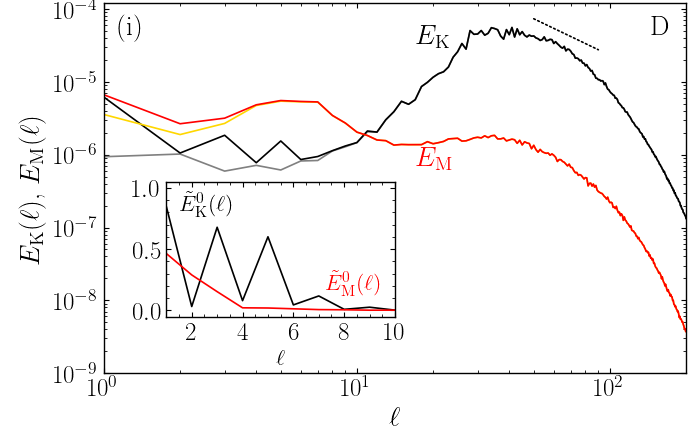}
\caption{Panels (a), (d), (g): Azimuthally averaged radial magnetic
  field $\mean{B}_r(\theta,t)$ from $r=R$ for runs A (top), C
  (middle), and D (bottom). Panels (b), (e), (h): Azimuthally averaged
  toroidal field $\mean{B}_\phi(\theta,t)$ from $r=0.65R$. The field
  strength is given in terms of the equipartition field $\Beq$. Panels
  (c), (f), (i): Power spectra of the velocity ($E_{\rm K}$) and
  magnetic fields ($E_{\rm M}$) from $r=0.85R$ as functions of
  spherical harmonic degree $\ell$. The grey (yellow) lines indicate
  non-axisymmetric ($m\neq0$) contributions. Dotted lines indicate the
  Kolmogorov $\ell^{-5/3}$ scaling. The inset shows the normalized
  fraction of the axisymmetric ($m=0$) contributions.}
\label{fig:fig1}
\end{figure*}

The azimuthally averaged radial magnetic fields near the surface of
the star, and the toroidal magnetic field near the base of the CZ for
runs A, C, and D are shown in Figure~\ref{fig:fig1} panels (a, d, g)
and (b, e, h), respectively. The rest of the runs follow very similar
patterns. All of these cases show a solar-like pattern of magnetic
field evolution at the surface with strong radial fields concentrated
in latitudes $|\theta|<50\degr$ and activity propagating
equatorward. A weaker poleward branch is visible in some runs; see
Figure~\ref{fig:fig1}(b). While a similar pattern can be seen for
$\mBBp$ near the surface, the dynamo wave propagates poleward near the
base of the CZ.

\begin{figure*}
\begin{interactive}{animation}{Bfieldt.mp4}
  \includegraphics[width=.95\textwidth]{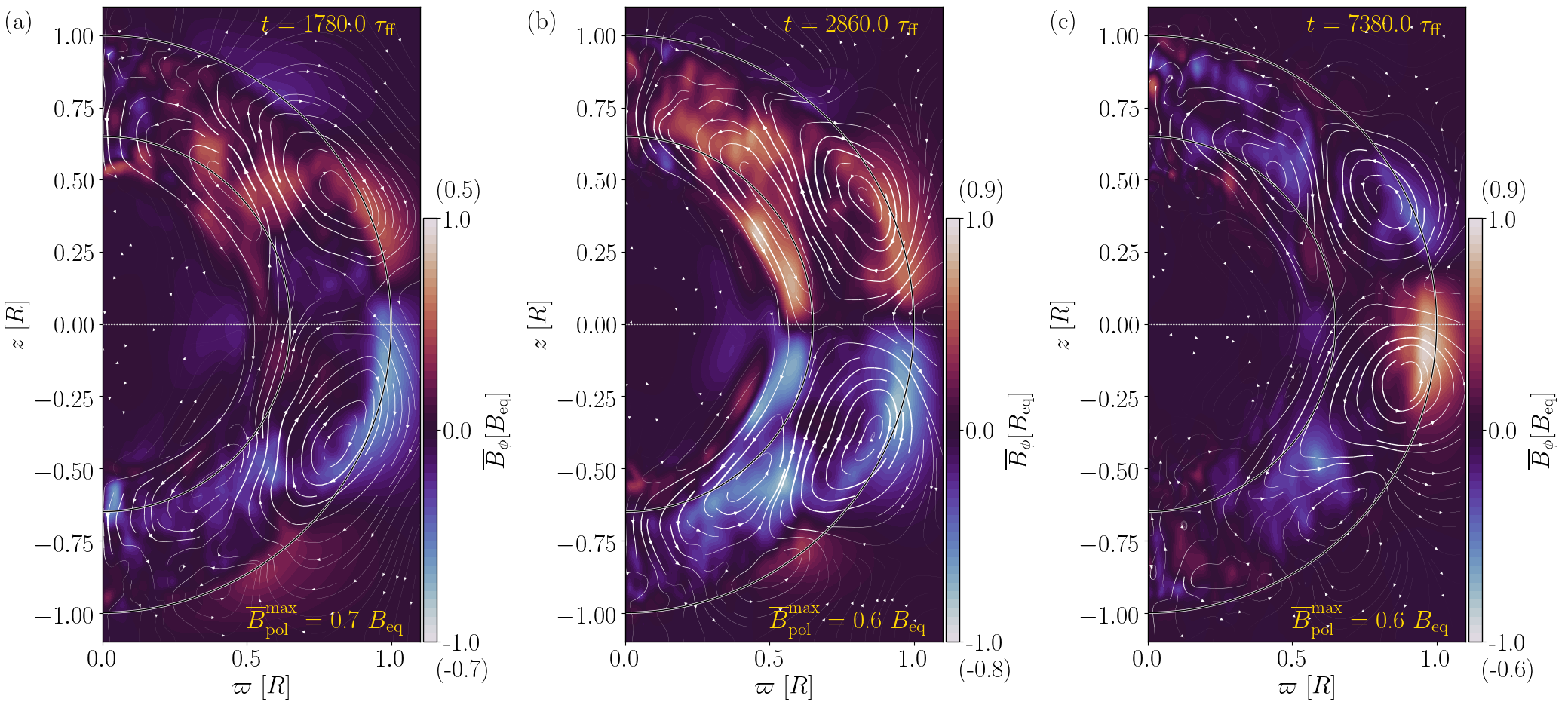}
\end{interactive}
\caption{Azimuthally averaged magnetic fields in units of $\Beq$ in
  cylindrical coordinates $(\varpi,z)$ for runs~A (a), C (b), and D
  (c) as functions of time for a time span of $3\times10^3$
  free-fall times. The colour contours show the toroidal fields and
  the arrows indicate poloidal fields. The grey lines indicate the
  approximate bottom of the CZ ($r=0.65R$) and the surface of the
  star ($r=R$).}
\label{fig:vid1}
\end{figure*}

Power spectra of the velocity and magnetic fields from a spherical
harmonic decomposition are shown in Figure~\ref{fig:fig1}(c, f, i) for
runs A, C, and D. The magnetic power spectrum has its maximum at the
largest possible scale ($\ell=1$) for all runs. The maximum of the
velocity power occurs at $\ell\gtrsim20$, and the peak moves to larger
$\ell$ with increasing rotation as the convective cells become
smaller. The spectra indicate a clear separation of scales between the
dominant scales of convection and those of the magnetic field. The
insets of Figure~\ref{fig:fig1}(c, f, i) show the fraction of the
power in the axisymmetric ($m=0$) part for $\ell\leq10$. In this
range, the equatorially asymmetric axissymmetric contributions with
odd $\ell$ dominate the velocity field. However, these contributions
are still clearly subdominant in the kinetic energy which is dominated
by convective flows with $\ell\gtrsim10$. The large-scale magnetic
fields are dominated by the axisymmetric $(\ell,m)=(1,0)$ component
with half or more of the total power in all cases.

The evolution of the mean magnetic fields in runs A, C, and D is shown
in the animations in \Fig{fig:vid1}. These visualizations show that
strong magnetic fields are concentrated near the surface of the star
outside the tangent cylinder, and near the interface between radiative
and convection zones inside the tangent cylinder. Strong magnetic
fields can also be found within the CZ at higher latitudes. This could
suggest the presence of multiple dynamo modes which have been detected
in simulations previously
\citep[e.g.][]{KKOBWKP16,2016ApJ...826..138B}. In distinction to these
studies, the cycles in the deep parts and near the surface in the
current simulations are synchronised such that their periods are the
same. Magnetic fields also penetrate into the upper part of the
radiative core down to a depth $r\approx0.5R$. The extent of this
penetration is highly likely unrealistic but such effects, albeit
quantitatively different, can still conceivably occur at the
interfaces of stellar radiative and convective zones.

\begin{figure*}
  \includegraphics[width=0.33\textwidth]{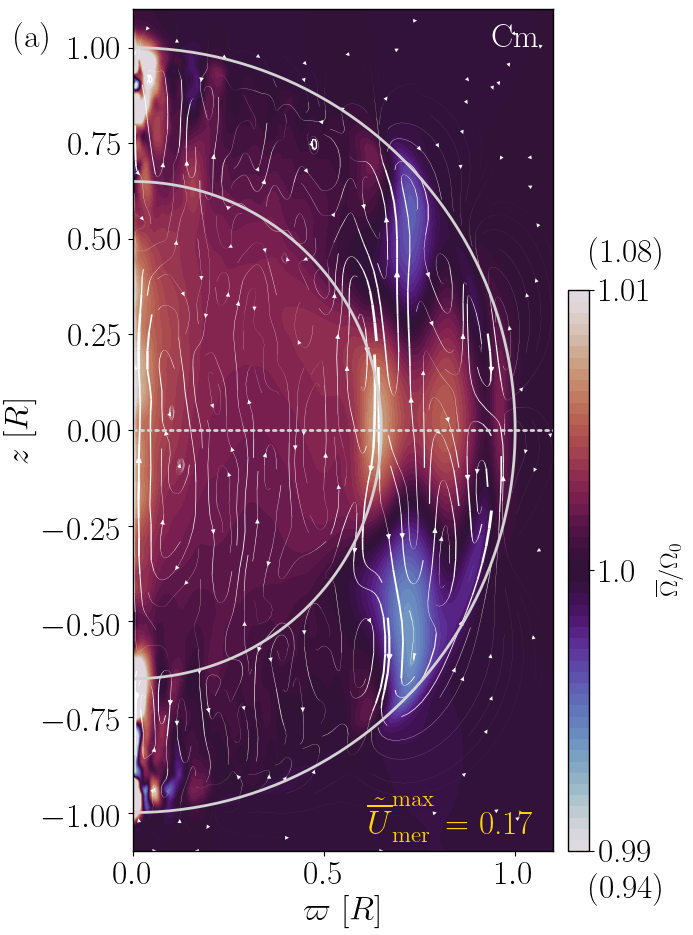}
  \includegraphics[width=0.33\textwidth]{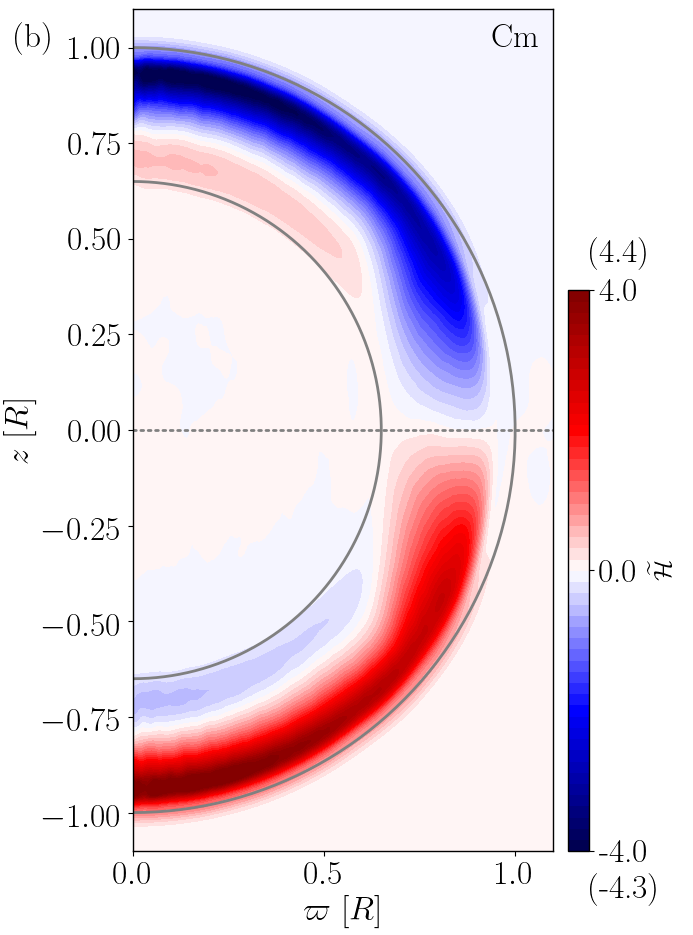}
  \includegraphics[width=0.33\textwidth]{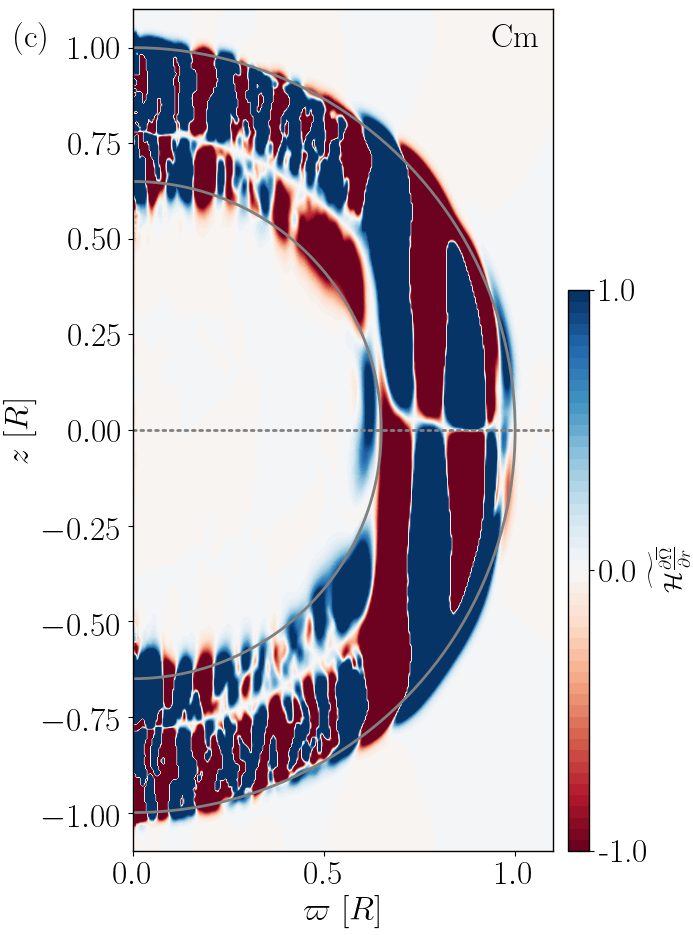}
  \caption{Normalized mean angular velocity
    $\mOm/\Omega_0=\mean{U}_\phi/(r\sin\theta\Omega_0)+1$ (colour
    contours) and meridional circulation (arrows) (a), normalized
    kinetic helicity $\tilde{{\cal H}} = {\cal H}/(\urms\orms)$ (b),
    and the normalized product $\widetilde{{\cal H} \frac{\pd
        \mOm}{\pd r}}$ (c) from run Cm.}
\label{fig:fig2}
\end{figure*}

\subsection{Flow states and dynamo considerations}

All of the current simulations are in the solar-like differential
rotation regime, which is characterised by a faster equator and slower
poles; see \Fig{fig:fig2}(a). However, the differential rotation is in
general weak, such that the amplitude is typically of the order of a
couple of per cent of $\Omega_0$ everywhere except near the axis where
the data is poorly converged. This is also apparent from the seventh
and eighth columns Table~\ref{tab:tab1} which show that the energies
associated with differential rotation and meridional circulation are
at most a few per cent of the total kinetic energy. This is partly due
to the damping of flows in the exterior which exerts a torque that
opposes differential rotation. Nevertheless, the current simulations
also show a minimum of $\mOm$ at mid-latitudes, which has been
conjectured to be the cause of equatorward migration by a dominating
$\alpha\Omega$ dynamo in earlier spherical shell simulations
\citep{WKKB14}.

The other ingredient in such models is the kinetic helicity, ${\cal
  H}=\mean{\ooo\bm\cdot\UUU}$, where $\ooo=\bm\nabla\times\UUU$ is the
vorticity, which is negative (positive) in the upper (lower) part of
the CZ in the northern hemisphere; \Fig{fig:fig2}(b). A sign change of
${\cal H}$ occurs in the deep layers of the CZ everywhere except near
the equator, which is a common feature in overshooting convection
\citep[e.g.][]{OSB01,KKB09a}. Thus the sign of kinetic helicity
follows the sign of $\gggg\bm\cdot\bm\Omega$ in the bulk of the
convection zone, and a significant helicity inversion and a consequent
reversal of the dynamo wave as suggested by \cite{DWBG16} does not
occur.

The product of the radial gradient of $\mOm$ and $\cal{H}$ determines
the propagation direction of the dynamo wave in $\alpha\Omega$ dynamos
\citep[][]{Pa55b,Yo75}. \Figu{fig:fig2}(c) indicates both poleward
(${\cal H}\pd_r\mOm<0$) and equatorward (${\cal H}\pd_r\mOm>0$)
regions outside the tangent cylinder, and predominantly poleward
propagation in the lower part of the CZ inside the tangent
cylinder. It is tempting to associate the equatorward branch near the
surface with a corresponding patch of positive $\widetilde{{\cal
    H}\frac{\pd\mOm}{\pd r}}$ and the poleward branch in the deep
parts with a corresponding patch of negative $\widetilde{{\cal H}
  \frac{\pd\mOm}{\pd r}}$ (in the northern hemisphere). However, the
Parker-Yoshimura rule applies strictly only in the case of a pure
$\alpha\Omega$ dynamo with spatially constant helicity and turbulent
diffusion. The low ratio of toroidal to poloidal magnetic energies
(10th and 11th columns in \Table{tab:tab1}) also suggests that the
dynamos in the current simulations are not of $\alpha\Omega$ type.

In addition to the already mentioned helicity inversion, further
possibilities to excite equatorward propagation include the
near-surface shear \citep[e.g.][]{Br05}, and an $\alpha^2$ dynamo with
a sign change of helicity at the equator
\citep[e.g.][]{MTKB10}. Neither of these possibilities can be ruled
out immediately with the data at hand, although the shear near the
surface is not very prominent in the current simulations. However,
explanations based on such simple models should be considered with
caution in light of a recent study by \cite{2021ApJ...919L..13W}, who
found that to explanation the cause and evolution of large-scale
magnetism in a spherical shell simulation required a mean-field model
with 24 turbulent transport coefficients derived using the test-field
method \citep[e.g.][]{SRSRC07}.

\begin{figure}
  \includegraphics[width=\columnwidth]{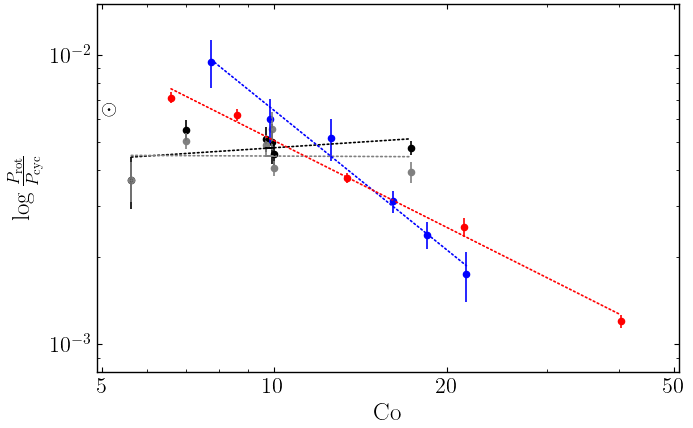}
  \caption{Rotation to cycle period as function of $\Co$ with
    $\Prot/P_{\rm cyc}^{\rm surf}$ (black), $\Prot/P_{\rm cyc}^{\rm
      deep}$ (grey). Additional data is from
    \cite{2018A&A...616A..72W} (red) and \cite{2018ApJ...863...35S}
    (blue). The dotted lines show linear fits to the data. Data point
    for the Sun ($\odot$) is indicated at the left $y$-axis.}
\label{fig:fig4}
\end{figure}

\subsection{Rotational scaling of dynamo cycles}

Long-term observations of chromospheric emission of late-type stars
suggest that many such stars have magnetic cycles similar to the Sun
\citep[e.g.][]{1995ApJ...438..269B}. It has also been suggested that
the ratio of stellar rotation and cycle periods fall into a number
branches as a function of the Coriolis number
\citep[e.g.][]{BST98,2017ApJ...845...79B}. These studies suggest
inactive and active branches with $\Prot/\Pcyc\propto\Co^\beta$ where
$\beta>0$. However, the exact nature \citep{2018A&A...619A...6O}, and
the signifigance of the branches continues to be debated
\citep[e.g.][]{2018A&A...616A.108B}. It is nevertheless interesting to
measure the ratio $\Prot/\Pcyc$ from simulations to see if any
systematics can be found. This has been done in a handful of studies:
\cite{2018A&A...616A..72W} found $\beta=-0.98\pm0.04$ whereas a
somewhat steeper relation with $\beta=-1.6\pm0.14$ was reported by
\cite{2018ApJ...863...35S}. Furthermore, the results from a more
heterogeneous set of simulations by \cite{2018A&A...616A.160V} also
suggest $\beta<0$. A notable exception is the study of
\cite{2019ApJ...880....6G} who found a flat ($\beta\approx0$) or
slightly positive $\beta$ for rapid rotation.

The cycle periods from the current simulations are computed using
\emph{libeemd} library \citep{Luukko2015} using the ensemble empirical
mode decomposition (EEMD). Periods are determined from
$\mBBr(R,\theta_d,t)$ and $\mBBp(0.65R,\theta_d,t)$, where
$-45\degr<\theta_d<45\degr$ is the range of latitudes considered. The
mode with the largest energy is identified as the primary cycle and
the period is computed from zero crossings of that mode. The mean
cycle period $P_{\rm cyc}$ is taken to be the average over
$\theta_d$. Error estimates are provided by dividing the time series
in three parts and repeating the analysis for each part. The largest
deviation from the mean period over the full time series is taken to
represent the error. The current results indicate that the ratio
$\Prot/\Pcyc$ is almost independent of $\Co$ in the parameter range
explored here with $\beta=0.13\pm0.17$ ($\beta=-0.01\pm0.21$) for the
surface (deep) cycles; see \Fig{fig:fig4}. These results differ
qualitatively from those of \cite{2018A&A...616A..72W} and
\cite{2018ApJ...863...35S} that are also shown in
\Fig{fig:fig4}\footnote{The Rossby numbers from
  \cite{2018ApJ...863...35S} were converted to Coriolis numbers as
  $\Co=2\pi/\Ro_{\rm b}$.}. \cite{2019ApJ...880....6G} reported
results for $\Prot/\Pcyc$ that are similar to those obtained here from
models that also included a radiative layer below the CZ. However, the
magnetic field configurations achieved in that study are quite
different from those here such that, e.g., no clear equatorward
migration is obtained.

The ratio $\Prot/\Pcyc$ in the current simulations is between
$(3\ldots5)\times10^{-3}$ which is close to the solar value
$\Prot^\odot/\Pcyc^\odot\approx6.5\times10^{-3}$ with
$\Prot^\odot=26$~days and $\Pcyc^\odot\approx11$~years. Coupled with
the near independence of $\Prot/\Pcyc$ on $\Co$ suggests that the
dynamos in the current simulations might capture some of the
characteristics of stars in the inactive branch where also the Sun
belongs to \citep[e.g.][]{BST98}. However, at the same time it is
clear that the current simulations do not reproduce many other aspects
of the Sun, such as the structure and magnitude of the differential
rotation.

\section{Conclusions} \label{sec:discussion}

Star-in-a-box simulations of a solar-like convective envelope were
shown to produce solar-like magnetic activity on a limited range of
rotation rates. The current simulations share many characteristics of
earlier spherical shells models \citep[e.g.][]{KMB12}, including a
local minimum of $\mOm$ at midlatitudes which has been conjectured to
be the cause of equatorward migration in those studies
\citep[][]{WKKB14}.  However, the magnetic cycles in the current
simulations show differences to the earlier studies in that the
equatorward migration of the active latitudes is not restricted to
mid-latitudes with negative radial differential rotation. Furthermore,
the rotational scaling of the cycles is qualitatively different from
the earlier studies in spherical shells
\citep[e.g.][]{2018A&A...616A..72W}, with a weak dependence of
$\Prot/\Pcyc$ on rotation.

However, several differences to earlier studies can be readily
identified. These include the addition of a simplified corona which
provides a free surface for the magnetic field rather than imposing
simplified surface boundary conditions \citep[see
  also][]{WKKB16}. Another difference is the inclusion of the
radiative core where strong magnetic fields can be stored and possibly
amplified by means other than helical convection
\citep[e.g.][]{2019ApJ...880....6G}. Finally, changing the geometry
and size of the system also allows, in general, a wider spectrum of
dynamo modes that can be excited. A more detailed analysis of the
maintenance of the magnetic fields is needed to precisely pinpoint the
differences to the earlier studies. Nevertheless, the current results
suggest that regions outside of the CZ shape the global dynamo
solutions, and that perhaps the dynamos in the Sun and other inactive
stars harbor dynamos where such effects are important.

\begin{acknowledgments}
This work was supported by the Deutsche Forshungsgemeinschaft
Heisenberg grant KA4825/4-1. The simulations were made using the
HLRN-IV supercomputers Emmy and Lise hosted by the North German
Supercomputing Alliance (HLRN) in G\"ottingen and Berlin, Germany.
\end{acknowledgments}

\software{Pencil Code \citep{2021JOSS....6.2807P}, 
          numpy \citep{harris2020array}
          }

\bibliography{bib_global}{}
\bibliographystyle{aasjournal}

\end{document}